\documentclass[11pt]{article}

\usepackage{graphicx}
\usepackage{amssymb}
\usepackage{amsmath}
\def\beq{\begin{equation}}
\def\eeq{\end{equation}}
\def\ba{\begin{array}}
\def\ea{\end{array}}
\def\bea{\begin{eqnarray}}
\def\eea{\end{eqnarray}}

\def\sq2{\sqrt{2}}

\def\ab1{{\rm ab}^{-1}}
\def\End{\end{document}}

\newcommand{\gm}{\gamma^\nu}
\newcommand{\smn}{\sigma^{\mu \nu}}
\newcommand{\gmc}{\gamma_\nu}
\newcommand{\smnc}{\sigma_{\mu \nu}}
\newcommand{\sqd}{\sqrt{2}}

\begin{document}

\title{Two-to-two processes at an electron-muon collider.}
\author{ Antonio O. Bouzas\thanks{abouzas@cinvestav.mx} \
  and F. Larios\thanks{francisco.larios@cinvestav.mx,
  corresponding author.} 
\\
Departamento de F\'{\i}sica Aplicada,
CINVESTAV-M\'erida, \\ A.P. 73, 97310 M\'erida,
Yucat\'an, M\'exico}


\maketitle

\begin{abstract}
    Based on a recent proposal to build an electron-muon collider, we
    study two-to-two production processes $e^-\mu^+\rightarrow
    f\bar{f}$, $\gamma\gamma$ that originate from dimension 6 and 8
    operators.  We compare the sensitivity to those effective
    couplings obtained at the collider with that of low energy
    measurements of $\mu\rightarrow e\gamma$, $\mu\rightarrow
    e\bar{e}e$ and $\mu\rightarrow e$ conversion that have recently
    been reported in the literature.  Whereas for the production of
    first family fermions the sensitivity of the collider processes is
    much weaker, for the second and third familiy fermions it is similar or 
    stronger than that of low-energy processes.  In the case of
    $e^-\mu^+\rightarrow \gamma\gamma$, the sensitivity to a dimension
    8 contact operator turns out to be the strongest in comparison.
\end{abstract}

\section{Introduction}

With the main objective of measuring Higgs boson properties,
there has been renewed interest in muon colliders
\cite{muoncoll, muonproton, muonion}.
At the low energy level, there have been numerous works concerning
$(g-2)_\mu$ from the theoretical and experimental point of
view \cite{aoyama, banerjee, paradisi, abbiendi, carloni}.
In addition, the recent confirmation of a
significant deviation from the Standard Model (SM) value
of $(g-2)_\mu$ at Fermilab \cite{gm2} has also encouraged
the proposal of new physics models
\cite{muonphys}-\cite{muonphys8} as well as the calculation
of higher order QED contributions to the $e\mu\to e\mu$
scattering process \cite{bonciani, heller}.
The idea of electron-muon collisions has been analyzed
in past decades \cite{oldemus}-\cite{oldemus6},
and recently it has been brought up again in a study of
an extra $Z'$ boson with generic couplings that could be
searched for through a Lepton Flavor Violating (LFV)
process like $e^- \mu^+ \to e^+ \mu^-$ \cite{bossi}.
In a subsequent article, the construction of a high energy
$e^- \mu^+$ collider has been proposed \cite{lulevin}.
Besides the clear capability of probing muon LFV
effects like the $He\mu$ coupling, this machine could
even test the Higgs-bottom quark coupling \cite{lulevin}.
One important advantage of such a machine is the very
low level of SM background processes as they come mostly
from vector boson fusion. Other than the elastic
$e^- \mu^+ \to e^- \mu^+$ there is no two-to-two process
to contend with and this is what motivates our study.
We are interested in fermion pair and two-photon
production that can come from contact terms
$e^- \mu^+ f\bar f$ and $e^- \mu^+ \gamma \gamma$.

LFV operators are already being strongly constrained from low energy
measurements like muon decays and $\mu - e$ transitions
\cite{davidson1, davidson2, davidson3}.  We will show the comparison
of the potential limits estimated here with these precision
measurements.
For $e^- \mu^+ \to f\bar f$ we consider the four-fermion dimension-6
operators of the Standard Model Effective Field Theory (SMEFT)
as given by the well known {\it Warsaw} basis \cite{grz10}.
We will also consider some dimension 8 operators that have been
constrained in \cite{davidson2}.
Given the chiral structure of the SMEFT, it is
straightforward to obtain the corresponding amplitudes in the helicity
basis.  For the two-photon production $e^- \mu^+ \to \gamma \gamma$
we address the contribution from
the LFV dipole operator, that is very strongly constrained by
$\mu \to e \gamma$, and we consider as well a dimension 8 contact
operator that is bound more strictly by our results than by low-energy
measurements.

The paper is organized as follows.  In section \ref{sec:two} we
analyse the two-body processes induced by four-fermion $e\mu ff$
operators.  We explicity provide the corresponding helicity amplitudes
and cross sections. Then, we estimate limits on the Wilson coefficients
for 13 dimension 6 and 5 dimension 8 four-fermion operators.
In section \ref{sec:aa} we consider $\gamma\gamma$ production
originating from two different contributions:
a trivalent dimension 6 dipole and a dimension 8 contact operator.
We write down the sum of squared amplitudes and the total
cross sections.  We estimate limits on the coefficients in the
same way we did in the previous section, but then we carry out
a detailed Monte Carlo analysis of signal and background
processes.   In this way we show that our simplified strategy
to obtain limits is realistic.  Finally, in
section~\ref{conclusions} we summarize our results.


\section{Four fermion operators.}
\label{sec:two}

Since we are dealing with massless chiral fermions, the
amplitudes are most conveniently written in the helicity basis
(see for example Ref.~\cite{mangano}).
Assuming some degree of polarization in the incoming beams,
the general form of the cross section can be divided
in four terms:
\bea
\sigma_{Pe^- P\mu^+} &=&  \frac{1+\mathcal{P}_{e^-}}{2}
\frac{1+\mathcal{P}_{\mu^+}}{2} \; \sigma_{++}
\; +\; \frac{1-\mathcal{P}_{e^-}}{2}
\frac{1-\mathcal{P}_{\mu^+}}{2} \; \sigma_{--}   \nonumber \\
&+& \frac{1+\mathcal{P}_{e^-}}{2}
\frac{1-\mathcal{P}_{\mu^+}}{2} \; \sigma_{+-}
\; +\; \frac{1-\mathcal{P}_{e^-}}{2}
\frac{1+\mathcal{P}_{\mu^+}}{2} \; \sigma_{-+} \;.
\label{polarcross}
\eea

In this section we will address four fermion contact operators
and we will obtain sensitivity limits with the following method:
first, for each operator we will assume a total polarization
in the cross section.  For instance, in
$e_R^- \mu_R^+ \to u_R {\bar u}_R$ that comes from the
right-chirality $Q_{eu}$ operator the ${\cal M} (+-+-)$ amplitude
gives rise to the $\sigma_{+-}$ cross section term, and we assume
$\mathcal{P}_{e^-}=+1$, $\mathcal{P}_{\mu^+}=-1$.
Second, we neglect backgrounds and obtain sensitivities in optimal
conditions.
Third, we will assume an integrated luminosity of order
$1$ ${\rm ab}^{-1}$ (as in Ref.~\cite{lulevin}) and require
a minimum cross section of $0.04 {\rm fb}$ from the operator
contribution.   Such a cross section yields a significant amount
of $40$ events.  In this way, since we are working with ideal
conditions we will avoid being too optimistic.  There are $13$
dimension 6 four-fermion operators and 9 of them only
contribute to one of the four polarized cross sections.
In their case, if there is only partial
beam polarization the actual number of events could be reduced
by about half or even by one quarter in case of unpolarized beams.
On the other hand, we do not expect the presence of background
to reduce the sensitivity significantly.  In the next section we will
corroborate that this is true for $\gamma \gamma$ production;
where we perform a full signal vs background analysis.

\subsection{Helicity amplitudes.}

In this work we are interested in the process
$e^-(p_1) \mu^+(p_2) \to f(p_3) {\bar f}(p_4)$
that does not exist in the SM at tree level, but that is generated
by dimension six four-fermion operators in the SMEFT.  Hereafter,
$f\bar f$ will stand for any of the three charged leptons $e$,
$\mu$ and $\tau$, or any of the quarks except the top quark.
The list of four fermion effective operators in Ref.~\cite{grz10} 
include arbitrary $(p,r,s,t)$ flavor indices to take into account.
However, this does not mean that there are so many possible
different helicity amplitudes as there are also Fierz
identities that relate them.    For instance, the operator
$O_{ll} = \bar l_{p} \gmc l_{r}  \bar l_{s} \gm l_{t}$
gives rise to the same amplitude
$e^- \mu^+ \to e^- e^+$ for any
combination $2111$, $1211$, $1121$ and $1112$.  We choose to
work with $2111$ that we denote as
$O^{2111}_{ll} = \bar l_{\mu} \gmc l_{e}  \bar l_{e} \gm l_{e}$.
For the processes $e^- \mu^+ \to \mu^- \mu^+$ and
$e^- \mu^+ \to \tau^- \tau^+$ the amplitudes from $O^{2122}_{ll}$
and $O^{2133}_{ll}$ are clearly the same.
In general, we shall take the flavor indices as $2111$ for
all the operators.   This specific choice may exclude some other
non-equivalent combinations in some operators, but we have found
that this assignment, in the end,
covers all the possible helicity amplitudes.
The purpose of this preliminary study, rather than being
comprehensive, is to get a first glimpse of the potential
sensitivity of the $e^- \mu^+$ collider in two-to-two processes
generated by contact operators
and compare with the sensitivities of low energy measurements.

The helicity amplitudes are shown in Table~\ref{tab:amps1}.
We see that $(\bar L L)(\bar L L)$ operators $Q_{ll}$,
$Q^{(1)}_{lq}$ and $Q^{(3)}_{lq}$ generate exactly the same
helicity amplitudes as they all involve left-chiral spinors.
On the other hand, the $(\bar L L)(\bar R R)$ operator
$Q_{ll} = \bar l_{\mu} \gamma_\mu l_{e} \bar e_{e} \gamma^\mu e_{e}$,
gives rise to two possible combinations of final state
chiralities: $e^-_R e^+_R$ and $e^-_R e^+_L$.  They are not
equivalent as shown in Table~\ref{tab:amps1}.
Notice that many helicity amplitudes are actually equal up to
a phase factor related to azimuthal angles.   The phase factor
is important only in the case of more than one diagram due to
interference effects.  We also point out that as a preliminary
analysis, we will be considering the contribution of each
operator separately.
Because of the relations \cite{mangano}:
\bea
\langle ij \rangle = -\langle ji\rangle \;, \;\;
 [ij]=-[ji] \; , \;\;
 \langle ij \rangle = [ji]^* \;, \;\;
  |\langle ij \rangle |^2 = 2 p_i \cdot p_j \; ,
\eea
there are really only three different amplitude structures
in this study: $[12][34]$, $[13][24]$ and $[14][23]$.
Their squares are proportional to the Mandelstam $s^2$,
$t^2$ and $u^2$ respectively.  So, they are actually not
independent.  In fact, there is an identity that is easy to
verify \cite{mangano}: $[12][34]=[14][32]+[13][24]$.

\begin{table}[ht!]
  \centering
  \begin{tabular}{|c|c|c|c|}\hline
    & ($\bar L L$)($\bar L L$)  &
$e^-(p_1) \mu^+(p_2) \to f(p_3) {\bar f}(p_4)$ & ${\cal M}$ 
    \\\hline
    $O_{ll}$ &
 $ \bar l_{\mu} \gmc l_{e}  \bar l_{e} \gm l_{e}$ &
    $e^-_{L} \mu^+_{L} \to e^-_{L} e^+_{L}$ &
    $2 [14] \langle 32\rangle$ 
    \\\hline
   $O_{lq}^{(1)}$ & $\bar l_{\mu} \gmc l_{e}  \bar q \gm q$ &
$e^-_{L} \mu^+_{L} \to u_{L} {\bar u}_{L}$ & $2 [14] \langle 32\rangle$ 
  \\\hline
$O_{lq}^{(3)}$ & $\bar l_{\mu} \gmc \tau^I l_{e} \bar q \gm \tau^I q$ &
$e^-_{L} \mu^+_{L} \to u_{L} {\bar u}_{L}$, $d_{L} {\bar d}_{L}$ &
  $\mp 2 [14] \langle 32\rangle$ 
  \\\hline
  & ($\bar R R$)($\bar R R$)  &
$e^-(p_1) \mu^+(p_2) \to f(p_3) {\bar f}(p_4)$ & ${\cal M}$ 
    \\\hline
    $O_{ee}^{}$ &
 $\bar e_{\mu} \gmc e_{e} \bar e_{e} \gm e_{e}$ &
    $e^-_{R} \mu^+_{R} \to e^-_{R} e^+_{R}$ & $2 \langle 14\rangle [32]$ 
     \\\hline
    $O_{eu}^{}$ &
 $ \bar e_{\mu} \gmc e_{e}  \bar u \gm u$ &
 $e^-_{R} \mu^+_{R} \to u_{R} {\bar u}_{R}$ &  $2 \langle 14\rangle [32]$
        \\\hline
    $O_{ed}^{}$ &
 $ \bar e_{\mu} \gmc e_{e}  \bar d \gm d$ &
 $e^-_{R} \mu^+_{R} \to d_{R} {\bar d}_{R}$ &  $2 \langle 14\rangle [32]$
   \\\hline
   & ($\bar L L$)($\bar R R$)  &
$e^-(p_1) \mu^+(p_2) \to f(p_3) {\bar f}(p_4)$ & ${\cal M}$
    \\\hline
    $O_{le}^{}$ &
 $ \bar l_{\mu} \gmc l_{e}  \bar e_{e} \gm e_{e}$ &
 $e^-_{L} \mu^+_{L} \to e^-_{R} e^+_{R}$  &  $2\langle 24\rangle [31]$ 
   \\\hline
 $O_{le}^{}$ &
 $ \bar l_{\mu} \gmc l_{e}  \bar e_{e} \gm e_{e}$ &
 $e^-_{R} \mu^+_{L} \to e^-_{R} e^+_{L}$ &  $2\langle 21\rangle [34]$ 
     \\\hline
   $O_{lu}^{}$ &
 $ \bar l_{\mu} \gmc l_{e}  \bar u \gm u$ &
 $e^-_{L} \mu^+_{L} \to u_{R} {\bar u}_{R}$ & $2 \langle 24\rangle [31]$ 
   \\\hline
    $O_{ld}^{}$ &
 $ \bar l_{\mu} \gmc l_{e}  \bar d \gm d$ &
   $e^-_{L} \mu^+_{L} \to d_{R} {\bar d}_{R}$ &  $2 \langle 24\rangle [31]$ 
    \\\hline
    $O_{qe}^{}$ &
 $ \bar q \gmc q  \bar e_{\mu} \gm e_{e}$ &
    $e^-_{R} \mu^+_{R} \to u_{L} {\bar u}_{L},d_{L} {\bar d}_{L}$ &
    $2 [24]\langle 31\rangle$ 
   \\\hline
   & ($\bar L R$)($\bar R L$)  &
$e^-(p_1) \mu^+(p_2) \to f(p_3) {\bar f}(p_4)$ & ${\cal M}$
    \\\hline
    $O_{ledq}^{}$ &
 $ \bar l_{\mu} e_{e}  \bar d q$ &
  $e^-_{R} \mu^+_{L} \to d_{R} {\bar d}_{L}$ &  $[12] [43]$ 
       \\\hline
       & ($\bar L R$)($\bar L R$)  &
$e^-(p_1) \mu^+(p_2) \to f(p_3) {\bar f}(p_4)$ & ${\cal M}$ 
    \\\hline
    $O_{lequ}^{(1)}$ &
 $ \bar l^j_{\mu} e_{e} \epsilon_{jk} \bar q^k u$ &
 $e^-_{R} \mu^+_{L} \to u_{R} {\bar u}_{L}$ &  $\langle 12\rangle [34]$ 
  \\\hline
    $O_{lequ}^{(3)}$ &
$\bar l^j_{\mu} \smnc e_{e} \epsilon_{jk} \bar q^k \smn u$ &
 $e^-_{R} \mu^+_{L} \to u_{R} {\bar u}_{L}$ & $4([13][24] + [14][23])$ 
  \\\hline
  \end{tabular}                               
  \caption{Dimension 6 operators with common flavor indices $(2111)$
and the helicity amplitudes generated.
    The $\frac{C_O}{\Lambda^2}$ factor to be included.  The $R,L$
    indices refer to chiralities. Final state $e^- e^+$ also stands for
    $\mu^- \mu^+$ and $\tau^- \tau^+$.  Similarly, $d\bar d$ also stands
    for $s\bar s$ and $b\bar b$; and $u\bar u$ also stands for $c\bar c$.
    Top quarks excluded.}
  \label{tab:amps1}
\end{table}

Now let us turn our attention to dimension 8 operators.  There are
multiple structures but we will pay attention to those specific
operators that have been bounded from $\mu \to e$ processes
\cite{davidson1,davidson2}.  Moreover, there are dimension 8 operators
that coincide with some dimension 6 operator except for an additional
$H^\dagger H$ term.  Obviously, their amplitudes would be equal except
for some rescaling factor.  The amplitudes of the dimension 8
operators that give rise to chiral structures that do not appear at
dimension 6 are shown in Table~\ref{tab:amps2}.

\begin{table}[ht!]
  \centering
  \begin{tabular}{|c|c|c|c|}\hline
    & ($\bar L R$)($\bar L R$)  &
 $e^-(p_1) \mu^+(p_2) \to f(p_3) {\bar f}(p_4)$ &  ${\cal M}$ 
    \\\hline
 $O_{le}^{(8)}$ & $ \bar l_{\mu} H e_{e}  \bar l_{e} H e_{e}$ &
    $e^-_{R} \mu^+_{L} \to e^-_{R} e^+_{L}$ &
    $\langle 12\rangle [43]$  
   \\\hline
 $O_{Tle}^{(8)}$ & $\bar l_{\mu} H\smn e_{e} \bar l_{e} H\smnc e_{e}$ &
   $e^-_{R} \mu^+_{L} \to e^-_{R} e^+_{L}$ &
   $4([31][24]+[41][23])$
   \\\hline
   $O_{leqd1}^{(8)}$ &
   $ \bar l_{\mu} H e_{e}  \bar q H d$ &
   $e^-_{R} \mu^+_{L} \to d_{R} {\bar d}_{L}$ &
   $\langle 12\rangle [43]$
   \\\hline
     $O_{leqd3}^{(8)}$ &
     $ \bar l_{\mu} H \smn e_{e}  \bar q H\smnc d$ &
   $e^-_{R} \mu^+_{L} \to d_{R} {\bar d}_{L}$ &
   $4([31][24] + [41][23])$ 
   \\\hline
    $O_{le}^{(8)}$ &
 $ \bar l_{\mu} H e_{e}  \bar l_{e} H e_{e}$ &
 $e^-_{R} \mu^+_{L} \to e^-_{L} e^+_{R}$ & $\langle 24\rangle [31]$  
    \\\hline
    $O_{Tle}^{(8)}$ &
 $\bar l_{\mu} H\smn e_{e} \bar l_{e} H\smnc e_{e}$ &
  $e^-_{R} \mu^+_{L} \to e^-_{L} e^+_{R}$ &  $4([21][34]+[14][23])$
    \\\hline    
     & ($\bar L R$)($\bar R L$)  &
 $e^-(p_1) \mu^+(p_2) \to f(p_3) {\bar f}(p_4)$  &  ${\cal M}$ 
    \\\hline
    $O_{leuq}^{(8)}$ &
 $ \bar l_{\mu} H e_{e}  \bar u {\tilde H}^\dagger q$ &
 $e^-_{R} \mu^+_{L} \to u_{R} {\bar u}_{L}$ &
    $\langle 12\rangle \langle 43\rangle$
       \\\hline
  \end{tabular}                               
  \caption{Helicity amplitudes generated by dimension 8 operators
    with common flavor indices $(2111)$ associated to four-fermion
    contact vertices that do not appear with dimension 6 operators.
A $\frac{v^2}{2} \frac{C_O}{\Lambda^4}$ factor is to be included.
As in Table~(\ref{tab:amps1}), final state $e^- e^+$ also stands for
$\mu^- \mu^+$ and $\tau^- \tau^+$; $d\bar d$ also stands for
$s\bar s$ and $b\bar b$; and $u\bar u$ also stands for $c\bar c$.}
  \label{tab:amps2}
\end{table}

\subsection{Cross sections and limits on coefficients.}

As mentioned above, there are only three different types of
helicity amplitudes squared.  Each of them gives rise to one
specific expression for the cross section.
For $|[12][34]|^2 =s^2$, after dividing
by the energy scale $\Lambda^4$ and integrating over
the phase space we obtain:  $\sigma_{1234} = s/(16\pi \Lambda^4)$.
For the center-of-mass (CM) collider energy $\sqrt{s} = 1.095$ TeV
($E_e=100,\;E_\mu=3000$GeV) we obtain
$\sigma_{1234} (\sqrt{s} = 1.095 {\rm TeV}) = 36.3$fb.  We will show
the corresponding cross sections for each operator in terms of this
common $\sigma_{1234}$.  The other amplitudes yield
$\sigma_{1324} = \sigma_{1423} = \sigma_{1234}/3$.  The amplitudes
obtained here grow with the collision energy, but our effective theory
calculation is valid for energies below the cut-off scale
$\Lambda =4$TeV.

Three benchmark collision energies are proposed in \cite{lulevin}:
(1) $E_e=20,\;E_\mu=200$GeV; (2) $E_e=50,\;E_\mu=1000$GeV; and
(3) $E_e=100,\;E_\mu=3000$GeV that correspond to CM energies
of $\sqrt{s}= 2\sqrt{E_e E_\mu} =\; 0.126 ,\; 0.447,\; 1.095$TeV,
respectively.  Since all cross sections are proportional to
$s|C_O|^2$ computing the bound at an energy $\sqrt{s_2}$
assuming we know the bound at an energy $\sqrt{s_1}$ is
straigthforward:  we just multiply by the ratio
$\sqrt{s_1}/ \sqrt{s_2}$.  For instance, $0.447/1.095 = 0.4$
and so we see that the limits at benchmark (3) will be more
than twice stronger than benchmark (2).  We shall focus on
the $E_e=100,\;E_\mu=3000$GeV benchmark in this study.  However,
in section \ref{sec:aa} we will see that the cross sections
for $e^- \mu^+ \to \gamma \gamma$ are not proportional to $s$,
but one is constant in energy and the other is proportional
to $s^2$.  We will provide limits obtained with the benchmark
(3) $E_e=100,\;E_\mu=3000$GeV and for greater energies.

The dimension 6 operators of Table~\ref{tab:amps1}
yield the following cross sections:
\bea
\frac{\sigma_{++}}{\sigma_{1234}} &=&
4|C_{le}^{}|^2 + N_c|C_{ledq}^{}|^2+N_c|C_{lequ}^{(1)}|^2
+\frac{16}{3} N_c|C_{lequ}^{(3)}|^2
\nonumber \\
\frac{\sigma_{+-}}{\sigma_{1234}} &=&
\frac{4}{3} |C_{ee}^{}|^2 + \frac{4}{3} N_c|C_{eu}^{}|^2 +
\frac{4}{3} N_c|C_{ed}^{}|^2 + \frac{4}{3} N_c|C_{qe}^{}|^2
\label{eq:cross6} \\
\frac{\sigma_{-+}}{\sigma_{1234}} &=&
\frac{4}{3} |C_{ll}^{}|^2 + \frac{4}{3} |C_{le}^{}|^2 +
\frac{4}{3} N_c|C_{lq}^{(1)}|^2 + \frac{4}{3} N_c|C_{lq}^{(3)}|^2
+\frac{4}{3} N_c|C_{lu}|^2 +\frac{4}{3} N_c|C_{ld}|^2 \;.
\nonumber
\eea
Where the $\sigma_{--}$ term does not appear for the flavor
assignment $2111$, but the operators that generate $\sigma_{++}$
would also generate $\sigma_{--}$ with the assignment $1211$.

By requiring that the value of a $C_Q$ coefficient be enough to
yield the minimum $0.04$fb of production cross section, we obtain
the following lower limits for $E_e=100,\;E_\mu=3000$GeV: 
\bea
C_{ll} \; ,\; C_{ee} \; &\geq& \; 2.88 \times 10^{-2}
\nonumber \\
C_{ledq} \; , C^{(1)}_{lequ} \; 
&\geq& \; 1.92 \times 10^{-2}
\nonumber \\
C_{le} \; ,\; C_{lu} \; , C_{ld} \; ,\; C_{qe} \;
&\geq& \; 1.66 \times 10^{-2}
\label{eq:limits6} \\
C_{eu} \; ,\; C_{ed} \; , C^{(1)}_{lq} \; ,
\; C^{(3)}_{lq} \;
&\geq& \; 1.66 \times 10^{-2}
\nonumber \\
C^{(3)}_{lequ} \; &\geq& \; 0.83 \times 10^{-2}
\nonumber
\eea

We can compare with the limits from low energy processes
$\mu \to e \gamma$, $\mu \to e\bar e e$ and $\mu \to e$
conversion in nuclei $\mu A \to e A$ as recently reported
in Tables 6 and 7 of Ref.~\cite{davidson1}.
We would like to point out that in almost all cases the
most stringent bounds in those tables come from $\mu A \to e A$
conversion in nuclei \cite{SINDRUMII}.  The exceptions are first
family four-lepton coefficients $C_{ll},C_{le},C_{ee}$ that are
constrained by $\mu \to e{\bar e}e$ (Table 6\cite{davidson1})
and $C^{(3)e\mu cc}_{lequ}$ that is constrained by
$\mu \to e\gamma$ (Table 7\cite{davidson1}).
Moreover, these $\mu A \to e A$ bounds are in fact
around four orders of magnitude stronger than the ones from
$\mu \to e \gamma$ and $\mu \to e\bar e e$.
This brings up one important observation: that if we ignored
$\mu A \to e A$ the conclusion would be that the $e\mu$
collider would yield much stronger constraints than any of
the low energy measurements.
In Table~\ref{tab:limits1} we show the limits to each operator
coefficient (at the scale $m_W$) and the ratio
$C^{\rm Low \; energy}_{Q}/C^{\rm Collider}_{Q}$ for each possible
final state.
Not surprisingly, we can observe that for the first family
fermions the limits are very stringent.  However,
for the second and third family modes that get their low
energy limits via loop contributions, the bounds are
weaker and about the same order of magnitude as the
potential limits from the $e^- \mu^+ \to f\bar f$
production.
Strictly speaking, the collider sensitivity for each
coefficient is taken at a scale of order 1 TeV; about ten
times higher than the electroweak scale.  In a more
precise analysis one should take into account the
renormalization group dependence on the scale.
However, one should bear in mind that such corrections
are usually of order a few percent.  For instance, the
coefficient $C_{DL}$ associated to the operator:
$(C_{DL} m_\mu/m_t^2) {\bar e}_R \smn \mu_L F^{\mu \nu}$.
An upper limit from its contribution to $\mu \to e\gamma$
decay is reported as $|C_{DL}|< 1.05 \times 10^{-8}$ at the
muon mass scale \cite{davidson1}.  The same limit becomes
$|C_{DL}|< 1.12 \times 10^{-8}$ at the $m_W$ scale,
which is only a $7\%$ numerical variation. 
For another example, let us consider the dimension 6
coefficient $C_{el}$ above.  In a series of articles
\cite{manohar,manohar2,manohar3,manohar4} we can
find the renormalization group evolution of the dimension
6 SMEFT operators listed in the Warsaw basis.
The running of $C_{el}$ in general depends on several other
operators, but let us take the term proportional to $C_{el}$
itself.  Then, we find that
$C_{le}(1TeV) \sim 0.96 C_{le} (m_W)$, that is only a
$4\%$ correction.  As we are only interested in a preliminary
assessment of sensitivities and comparisons with low energy
experiment limits we shall not take corrections of scale
dependence into account.
Recently, limits on the $C^{(1)}_{lq}$ and $C^{(3)}_{lq}$
coefficients have been published based on the LHC
$pp\to e\mu$ dilepton production that are approximately
$C^{(1,3)}_{lq} < 0.3$ for first family quarks,
$C^{(1,3)}_{lq} < 2.0$ for second family and
$C^{(1,3)}_{lq} < 5.0$ for $b\bar b$ \cite{angelescu}.
They project that with a hundred times more luminosity
the HL-LHC could reduce these limits to one third
of the current value.
From what we have found here, the $e\mu$ collider
would have at least one order of magnitude greater
sensitivity than the HL-LHC for these operators.

\begin{table}[ht!]
  \centering
  \begin{tabular}{|c|c|c|c|}\hline
 $C_{Q}$   & $C^{\rm Low}_{Q}$ ($C^{\rm Low}/C^{\rm Coll}$)  & 
    $C^{\rm Low}_{Q}$ ($C^{\rm Low}/C^{\rm Coll}$)  &
    $C^{\rm Low}_{Q}$ ($C^{\rm Low}/C^{\rm Coll}$)  
   \\
  & $e^- e^+$ & $\mu^- \mu^+$ & $\tau^- \tau^+$ 
  \\\hline
    $C_{ll}^{}$ &
    $4.16 \times 10^{-4}$ ($1.45 \times 10^{-2}$) &
    $0.98 \times 10^{-2}$ ($0.34$) &
  $1.97 \times 10^{-2}$ ($0.69$)

    \\\hline
    $C_{ee}^{}$ &
    $4.16 \times 10^{-4}$ ($1.45 \times 10^{-2}$) &
    $0.98 \times 10^{-2}$ ($0.34$) &
    $1.97 \times 10^{-2}$ ($0.69$)

      \\\hline
    $C_{le}^{}$ &
    $4.92 \times 10^{-4}$ ($2.96 \times 10^{-2}$) &
    $1.99 \times 10^{-2}$ ($1.20$) &
      $1.98 \times 10^{-2}$ ($1.19$)
 \\\hline
      
  & $d\bar d$ & $s\bar s$ & $b\bar b$ 
  \\\hline
    $C^{(1)}_{lq}$ &
    $1.51 \times 10^{-5}$ ($0.91 \times 10^{-3}$) &
    $1.60 \times 10^{-2}$ ($0.96$) &
    $2.49 \times 10^{-2}$ ($1.50$)

    \\\hline
    $C^{(3)}_{lq}$ &
    $2.69 \times 10^{-4}$ ($1.62 \times 10^{-2}$) &
    $6.08 \times 10^{-3}$ ($0.37$) &
    $2.49 \times 10^{-2}$ ($1.50$)
  \\\hline
    $C^{}_{ld}$ &
    $2.80 \times 10^{-5}$ ($1.69\times 10^{-3}$) &
    $1.97 \times 10^{-2}$ ($1.19$) &
    $2.49 \times 10^{-2}$ ($1.50$)
 \\\hline
    $C^{}_{ed}$ &
    $2.86 \times 10^{-5}$ ($1.72\times 10^{-3}$) &
    $1.97 \times 10^{-2}$ ($1.19$) &
    $2.49 \times 10^{-2}$ ($1.50$)
 \\\hline
    $C^{}_{qe}$ &
    $1.52 \times 10^{-5}$ ($0.92\times 10^{-3}$) &
    $1.58 \times 10^{-2}$ ($0.95$) &
    $2.50 \times 10^{-2}$ ($1.51$)

 \\\hline
    $C^{}_{ledq}$ &
    $5.34 \times 10^{-6}$ ($2.79\times 10^{-4}$) &
    $1.11 \times 10^{-4}$ ($5.80\times 10^{-3}$) &
    $3.66 \times 10^{-3}$ ($0.19$)

 \\\hline
 
  & $u\bar u$ & $c\bar c$ & $--$ 
  \\\hline
    $C^{}_{lu}$ &
    $3.30 \times 10^{-5}$ ($1.99 \times 10^{-3}$) &
    $0.88 \times 10^{-2}$ ($0.53$) &
 
  \\\hline
    $C^{}_{eu}$ &
    $3.19 \times 10^{-5}$ ($1.92 \times 10^{-3}$) &
    $0.89 \times 10^{-2}$ ($0.54$) &

  \\\hline
    $C^{(1)}_{lequ}$ &
    $5.45 \times 10^{-6}$ ($2.84 \times 10^{-4}$) &
    $0.97 \times 10^{-3}$ ($5.1\times 10^{-2}$) &

  \\\hline
    $C^{(3)}_{lequ}$ &
    $5.10 \times 10^{-5}$ ($6.1\times 10^{-3}$) &
    $5.34 \times 10^{-6}$ ($6.4\times 10^{-4}$) &
        
       \\\hline
  \end{tabular}                               
  \caption{Upper limits to the dimension 6 coefficients from
 the low energy experiments as reported in tables 6 and 7 of
 \cite{davidson1}.  They have been multiplied by a factor
 $(4/0.174)^2$ to account for the normalization scale $m_t$
 used by \cite{davidson1} instead of the $\Lambda = 4$TeV
 scale adopted in this study.  In addition, the ratio of
 limits Low/Collider is shown in parenthesis
 ($E_e=100,\;E_\mu=3000$GeV).}
  \label{tab:limits1}
\end{table}

The dimension 8 operators of Table~\ref{tab:amps2}
yield the following cross sections:
\bea
\frac{\sigma_{++}}{\sigma_{1234}} (e^-_R e^+_L) &=&
\frac{v^4}{4\Lambda^4}
\left( |C_{le}^{(8)}|^2 + N_c|C_{leqd1}^{(8)}|^2 +
N_c|C_{leuq}^{(8)}|^2 + \frac{16}{3} N_c|C_{leqd3}^{(8)}|^2
+\frac{16}{3} |C_{Tle}^{(8)}|^2
\right)  \nonumber \\
\frac{\sigma_{++}}{\sigma_{1234}} (e^-_L e^+_R) &=&
\frac{v^4}{4\Lambda^4}
\left( \frac{1}{3} |C_{le}^{(8)}|^2 +
\frac{112}{3} N_c|C_{Tle}^{(8)}|^2 \right)
\label{eq:cross8}
\eea
A comment on the tensor operators $Q^{(3)}_{lequ}$, $Q_{leqd3}$ and
$Q_{Tle}$ is in order here.  $Q^{(3)}_{lequ}$, $Q_{leqd3}$ and the
$\sigma_{++} (e^-_R e^+_L)$ contribution of $Q_{Tle}$ generate the
same amplitude $[13][24]+[14][23]$. Except for the $v^4/(4\Lambda^4)$
additional factor, the cross sections are equal.  However,
for the $\sigma_{+-} (e^-_L e^+_R)$ helicity mode the $Q_{Tle}$
amplitude is $[21][34]+[14][23]$.  In the CM
frame the functional dependence on the $p_3$ polar angle's
$\cos{\theta}$ is: $[13][24]+[14][23] \sim \cos{\theta}$.
If we compare with $[21][34]+[14][23] \sim (3+\cos{\theta})/2$
we can see that the latter yields a much greater cross section.

\begin{table}[ht!]
  \centering
  \begin{tabular}{|c|c|}\hline
 $C_{Q}$   & $C^{\rm Low}/C^{\rm Coll}$ 
  \\\hline
 $C_{le}^{(8)}$ & $5 \times 10^{-3} \; /14.0$  
  \\\hline
 $C_{Tle}^{(8)}$ & $2 \times 10^{-3} \; /1.3$  
  \\\hline
 $C_{leqd1}^{(8)}$ & $3 \times 10^{-5} \; /8.1$  
  \\\hline
 $C_{leqd3}^{(8)}$ & $7 \times 10^{-4} \; /3.5$  
  \\\hline
 $C_{leuq}^{(8)}$ & $3 \times 10^{-5} \; /8.1$  
       \\\hline
  \end{tabular}                               
  \caption{The ratio of potential limits from low energy processes and
minimum observable values at the collider for the dimension 8 coefficients
 (Table 5 of \cite{davidson2}; $E_e=100,\;E_\mu=3000$GeV).}
  \label{tab:limits2}
\end{table}

As in Table~\ref{tab:limits1}, in Table~\ref{tab:limits2}
we show the ratio $C^{\rm Low}/C^{\rm Coll}$ for the dimension
8 operators.  Comparing with the dimension 6 coefficients,
there is a suppresing $v^2/(2\Lambda^2)$ factor of order
$2\times 10^{-3}$ and the $C^{\rm Coll}$ minimum values have
to be much bigger.  In contrast, according to Table 5 of
Ref.~\cite{davidson2} the limits from low energy experiments
are still very stringent for dimension 8 couplings.

What we have learned from Table~\ref{tab:limits1} is that for
first family fermions $f\bar f$ the sensitivity of the low
energy measurement of $\mu A \to e A$ conversion in nuclei
is two or more orders of magnitude higher.   Maybe all the
operator coefficients are indeed very suppressed;
regardless of potential cancellations.  However, for most
of the second and third family $f\bar f$ states the
collider sensitivities are of the same order of magnitude
as the low energy limits. The electron-muon collider should
be able to provide additional and competitive limits to
constrain the set of dimension 6 four fermion LFV operators.
The same may not be true for dimension 8 operators, at least
for the ones that can be constrained by the low energy
experiments.


\section{The $e^- \mu^+ \to \gamma \gamma$ process.}
\label{sec:aa}

\begin{figure}[ht!]
\centering
\includegraphics[scale=0.4]{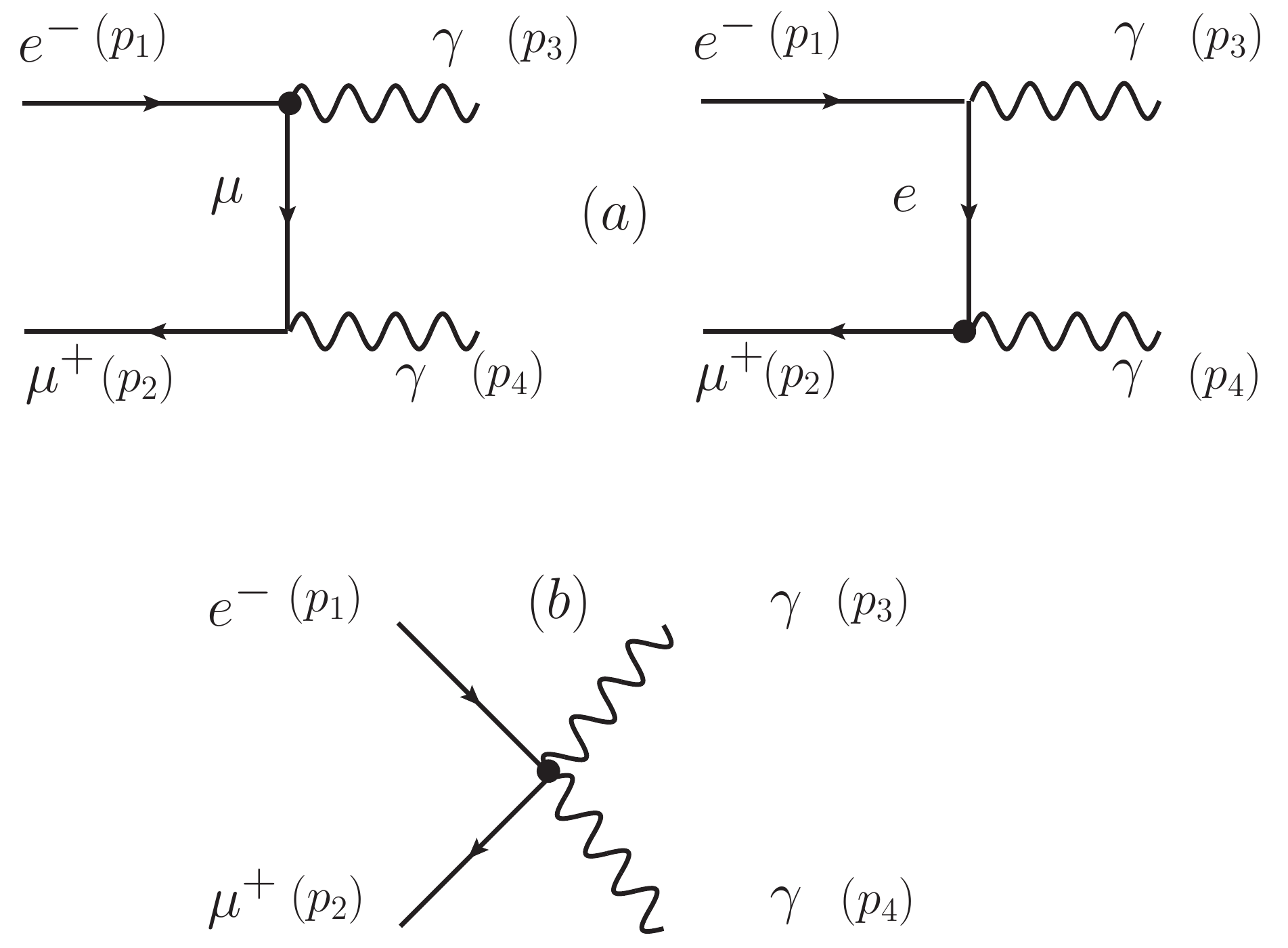}  
\caption{The anomalous two-to-two process
$e^- \mu^+ \to \gamma \gamma$.  Process (a) (t-channel)
as given by the trivalent vertex dipole operator $Q_{eA}$.
Process (b) as given by the contact term, dimension 8
operator $Q_{eAA}$.}
  \label{fig:emu}
\end{figure}

The $e^- \mu^+ \to \gamma \gamma$ amplitude can be generated
by the dimension 6 flavor-changing magnetic dipole operator
\bea
\label{eq:Q6}
Q_{eA} = \bar l_\mu \smn e_e H F_{\mu \nu} \; ,
\eea
through $t$ and $u$ channel diagrams where one of the
photons is emitted by the effective coupling.
Another possibility comes from the dimension 8 operator
\bea
\label{eq:Q8}
Q_{eAA} = \bar l_\mu e_e H F^{\mu \nu} F_{\mu \nu} \; .
\eea
We have two chiral versions: $Q_{eAL}$ and $Q_{eAR}$
($Q_{eAAL}$, $Q_{eAAR}$) referring to left-handed
and right-handed electron respectively.
In figure \ref{fig:emu} we show the Feynman diagrams
associated to each operator.

Both operators give rise to amplitudes that do not depend
on angles:
\bea
\sum |{\cal M}|^2 &=& \left( |C_{eAL}|^2 + |C_{eAR}|^2 \right)
e^2 \frac{v^2}{\Lambda^4} 48s \nonumber \\
&+& \left( |C_{eAAL}|^2 + |C_{eAAR}|^2 \right)
\frac{v^2}{\Lambda^8} 4s^3 \label{m2gaga}
\eea

They yield the total cross sections:
\bea
\sigma_{--} + \sigma_{++}
&=& \left( |C_{eAL}|^2 + |C_{eAR}|^2 \right)
\frac{3 e^2 v^2}{2 \pi \Lambda^4} \nonumber \\
&=& \left( |C_{eAL}|^2 + |C_{eAR}|^2 \right)
4.32 {\rm fb}\; .\label{sigmagaga} \\
\sigma_{--} + \sigma_{++}
&=& \left( |C_{eAAL}|^2 + |C_{eAAR}|^2 \right)
\frac{v^2}{8 \pi \Lambda^8} s^2 \nonumber \\
&=& \left( |C_{eAL}|^2 + |C_{eAR}|^2 \right)
0.0206 {\rm fb}\; ,\nonumber
\eea
where the numerical value on the second line is independent of the
collision energy, but the numerical value on the fourth line is taken
at $\sqrt{s}=1.095$ TeV. 

We see, then, that coefficients $C_{eAL(R)}$ and $C_{eAAL(R)}$ of
order $0.1$ and $\sqd$ respectively would give us a $\sigma =0.04$fb
value that is our minimum acceptable cross section.  Let us notice
that the current limit from $\mu \to e \gamma$ is of order
$5\times 10^{-6}$ for the dipole coefficient $C_{eAL(R)}$, well below
the potential collider sensitivity.  On the other hand, for the
dimension 8 operator coefficient $C_{eAAL(R)}$ the potential
sensitivity from $\mu A \to e A$ transitions is six orders of
magnitude less stringent: $C_{eAAL(R)}\leq 3.2$ \cite{davidson2}.  We
point out that, in this study, this last coefficient is the one with
the highest sensitivity by the $e\mu$ collider as compared to the low
energy measurements.

Seeing the relatively high sensitivity to the electromagnetic
$Q_{eAA}$ operator one may wonder what about the effective
vertex $e\mu GG$?  We can use the $Q_{eAA}$ calculation with
the analogous gluon operator 
\bea
Q_{eGG} = \frac{C_{eGG}}{\Lambda^4}
\bar l_\mu e_e H G^{a\mu \nu} G^a_{\mu \nu} \; .
\eea
The $e^- \mu^+ \to gg$ production the cross section is
now $N_c=8$ times greater than $e^- \mu^+ \to \gamma \gamma$
and a coefficient of order $0.5$ would yield the minimum
observable production.  However, according to \cite{davidson2}
the low energy $\mu A \to eA$ limit is three orders of
magnitude more stringent in this case:
$C_{eGG} \leq 1.6 \times 10^{-3}$.

As mentioned above, the amplitudes squared for
$e^- \mu^+ \to \gamma \gamma$ do not depend on the polar angle.
That means that in terms of rapidity, for instance $y=y^*_3$
in the CM frame:
\bea
\frac{d\sigma}{dy} = \frac{d\sigma}{dc_\theta}
\frac{dc_\theta}{dy} = a_{0}
\frac{4\exp(2y)}{(1+\exp(2y))^2} \; ,
\label{rapidityshape}
\eea
where $a_0$ is a constant.  The shape of the rapidity distribution
in the CM frame is then centered around zero with a width of
approximately $2$ units as shown below in figure
\ref{fig:distributions}, right column. In the lab frame the center
is shifted towards $-1.70$ (see next subsection).

\vspace{1cm}

\subsection{Monte Carlo analysis of $\gamma\gamma$ production and its
  SM background.}
\label{sec:numres.aa}

\begin{figure}[ht!]
\centering
\includegraphics[scale=0.35]{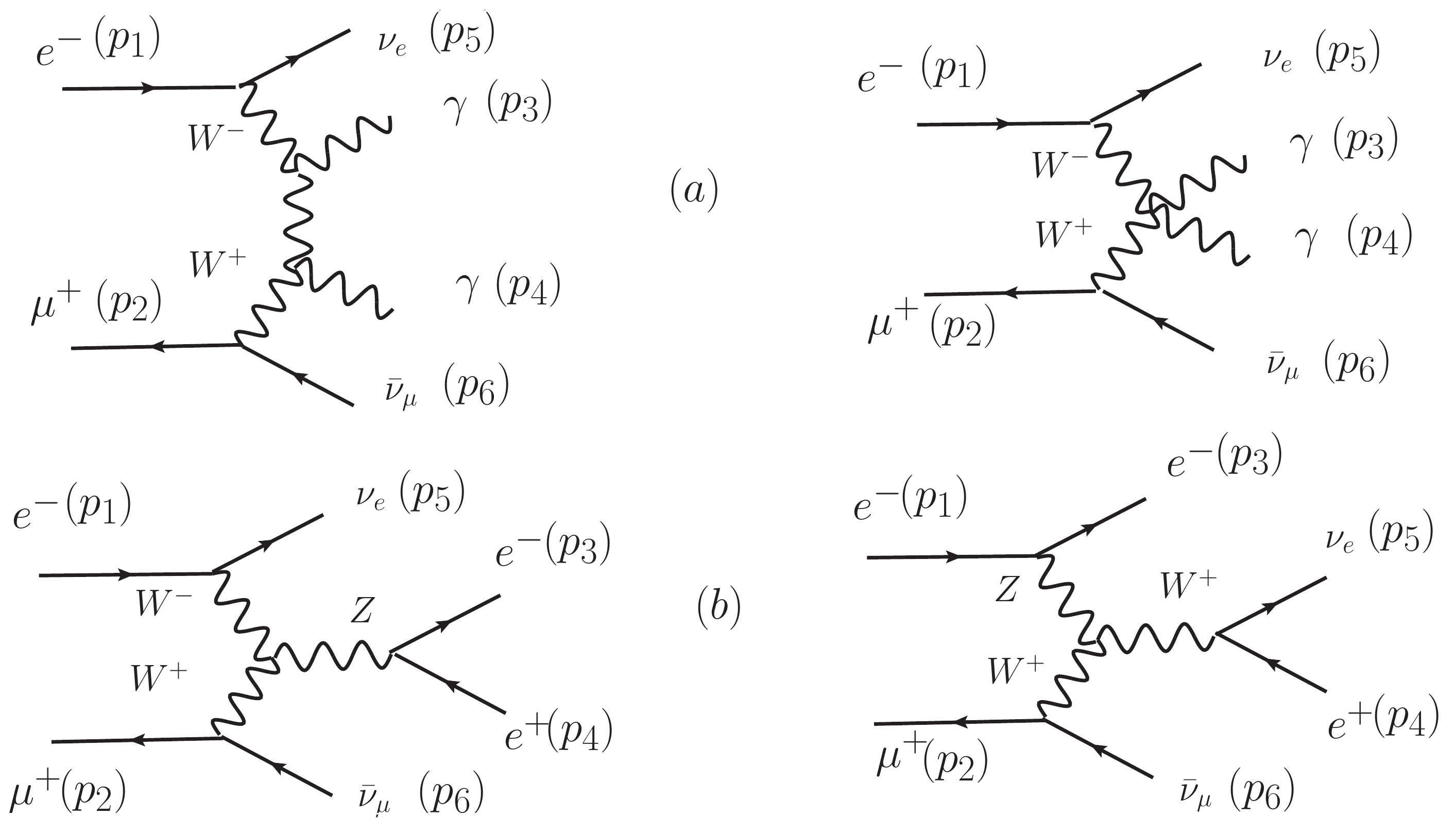}  
\caption{Representative unitary-gauge Feynman diagrams for the SM processes
  (a) $e^- \mu^+ \to \gamma \gamma \nu_e \bar \nu_\mu$ and
  (b) $e^- \mu^+ \to e^- e^+ \nu_e \bar \nu_\mu$.}
  \label{fig:emusm}
\end{figure}

In the SM, $\gamma \gamma$ production is given by the two-to-four
process $e^- \mu^+ \to \gamma \gamma \nu_e \bar {\nu_\mu}$, involving
13 Feynman diagrams in unitary gauge.  There is also $e^-e^+$
production $e^- \mu^+ \to \nu_e \bar {\nu_\mu} e^-e^+$, involving 24
Feynman diagrams.  In figure \ref{fig:emusm} we show two
representative diagrams for the $\gamma \gamma$ and resonant $e^- e^+$
production in the SM.
Signal and background have very different kinematics and this makes
the separation straightforward.  Our goal is to show how a basic set
of cuts can reduce the potential background dramatically.  We point
out here that in the lab frame the rapidities $y$ are shifted
with respect to those $y^*$ in the CM frame:
\bea
y = y^* - y_0 \; , \;\; {\rm with} \;\;
y_0=\frac{1}{2} \ln\left(\frac{E_\mu}{E_e}\right) = 1.70 \;,
\label{labtocm} 
\eea where $y_0=1.7$ is the shift value for $E_\mu/E_e =30$.  The muon
beam goes in the direction of $-\hat k$ and so the event products
usually appear on the backwards hemisphere.

Thus, to study $\gamma\gamma$ production in $e^-\mu^+$ collisions we
consider the three processes,
\begin{eqnarray}
  \label{eq:sign}
  e^- \mu^+  &\rightarrow& \gamma \gamma,\\
\label{eq:proc1}  
  e^- \mu^+  &\rightarrow& \gamma \gamma \bar{\nu}_\mu \nu_e,\\
  \label{eq:proc2}
    e^- \mu^+  &\rightarrow& e^- e^+ \bar{\nu}_\mu \nu_e,
\end{eqnarray}
where the first one is our signal process as depicted in figure
\ref{fig:emu} and (\ref{eq:proc1}), (\ref{eq:proc2}) the SM
backgrounds shown in figure \ref{fig:emusm}.  We implemented the
effective interactions (\ref{eq:Q6}), (\ref{eq:Q8}) in our Monte Carlo
simulations by means of \textsc{feynrules} 2.0 \cite{feynrules}.  We
simulated the signal and background processes with \textsc{madgraph}
2.6 \cite{madgraph}, with beam energies $(E_e, E_{\mu})=$ (100, 3000),
(150, 4500) and (200, 6000) GeV. We then have $\sqrt{s}=1.095$, 1.643
and 2.191 TeV, respectively.  In the two-body signal process
(\ref{eq:sign}), the CM energies of the final photons are fixed at
$E^*_3=E^*_4=\sqrt{s}/2$.  This is unlike what happens with the
backgrounds (\ref{eq:proc1}), (\ref{eq:proc2}), where there is a
continuous spectrum for $E^*_3$ and $E^*_4$.  Similarly, for the
signal process $|\vec{p}_{3T}+\vec{p}_{4T}|=0$, but for the
backgrounds $|\vec{p}_{3T}+\vec{p}_{4T}|=\not\!\!E_T$ which has a
continuous range of values.  Furthermore, we observe the final-state
photons in the signal process to be very central in the CM frame, with
$|y^*_{3,4}| = |y_{3,4}+y_0|\lesssim 2.5$ for the vast majority of
events, as expected in view of the analytical distribution
(\ref{rapidityshape}) and as shown below in figure
\ref{fig:distributions}.  For the process (\ref{eq:proc1}), without
restrictions on $E^*_{3,4}$, the photon rapidity distribution in the
lab frame is symmetric about $-y_0 = -1.70$, but very broad.  If we
require $E^*_{3,4}$ in (\ref{eq:proc1}) to be large, however, the
final-state photons must be very forward or backward.  For example, if
$E^*_{3,4}\simeq 250$GeV the rapidities will have maxima at
$-y_0\pm\Delta y$ with $\Delta y\simeq2$ as shown in figure
\ref{fig:distributions}; and $\Delta y$ gets larger for greater values
of $E^*_{3,4}$.  For the process (\ref{eq:proc2}) the electron
rapidity distribution presents essentially the same features, but is
less forward-backward symmetric, as seen in the figure.

We are, thus, led to consider the following set of phase-space cuts,
\begin{equation}
  \label{eq:cuts2}
  \begin{gathered}
  C_0:\; p_{3T}, p_{4T} > 1.0\,\mathrm{GeV},
  \qquad
  C_1:\; E_{3}^*, E_{4}^* > 500.0\,\mathrm{GeV},
  \\
  C_2:\; p_T^\mathrm{tot} = |\vec{p}_{3T}+\vec{p}_{4T}| <
  20.0\,\mathrm{GeV},
  \qquad
  C_3:\; |y_{3}+y_0|, |y_{4}+y_0| < 1.75.
  \end{gathered}  
\end{equation}
The cut $C_0$ is necessary to control infrared divergences in
(\ref{eq:proc1}), (\ref{eq:proc2}).  As discussed above, the cuts
$C_{1,2,3}$ in (\ref{eq:cuts2}) have only small effects on the signal
cross section, but they do substantially decrease the cross section
for the backgrounds.  The effect of the cuts (\ref{eq:cuts2}) on the
cross sections for the processes (\ref{eq:sign}), (\ref{eq:proc1}),
(\ref{eq:proc2}) is illustrated at $(E_e,E_\mu)=$ (100, 3000) GeV  in
table \ref{tab:sigmacuts}. 
\begin{table}[ht!]
  \centering
  \begin{tabular}{|c|c|c|c|c|c|c|}\hline
$\mathcal{P}_\mu$ & $\mathcal{P}_e$ & $\mathrm{cuts}$
 & $\sigma_{\gamma\gamma}^{(3)}[\mathrm{fb}]$
 & $\sigma_{\gamma\gamma}^{(4)}[\mathrm{fb}]$
 & $\sigma_{\nu\nu\gamma\gamma}[\mathrm{fb}]$
 & $\sigma_{\nu\nu ee}[\mathrm{fb}]$
  \\\hline
  $0.0$ & $0.0$ & $C_0$ &
  $2.16$ & $0.0103$ & $770.2$ & $426.6$
  \\\hline
  $0.0$ & $0.0$ & $C_{0,1}$ &
$2.16$ & $0.0103$ & $0.0049$ & $0.073$
  \\\hline
   $0.0$ & $0.0$ & $C_{0,1,2}$ &
$2.16$ & $0.0103$ & $0.0015$ & $0.0081$
    \\\hline
   $0.0$ & $0.0$ & $C_{0\mbox{--}3}$ &
$2.04$ & $0.0097$ & $3.23\times10^{-5}$ & $0.00108$
  \\\hline    
   $+0.4$ & $+0.8$ & $C_{0,1}$ &
  $2.86$ & $0.0136$ & $0.0014$ & $0.081$
  \\\hline
  $+0.4$ & $+0.8$ & $C_{0,1,2}$ &
  $2.69$ & $0.0128$ & $0.00043$ & $0.0082$ 
    \\\hline
  $+0.4$ & $+0.8$ & $C_{0\mbox{--}3}$ &
  $2.69$ & $0.0128$ & $9.85\times10^{-6}$ & $0.0038$
    \\\hline                                                                  
  \end{tabular}                               
  \caption{Cumulative effects of the cuts (\ref{eq:cuts2}) on the cross
    sections for (\ref{eq:sign}), (\ref{eq:proc1}), (\ref{eq:proc2})
    at $E_e=$ 100 GeV, $E_\mu=$ 3 TeV.}
  \label{tab:sigmacuts}
\end{table}
The cross section $\sigma_{\gamma\gamma}^{(3)}$ refers to the
anomalous process (\ref{eq:sign}) with only the trivalent $e\mu\gamma$
vertex and the Wilson coefficients $C_{eAL} = C_{eAR} =1$.
Similarly, $\sigma_{\gamma\gamma}^{(4)}$ refers to
(\ref{eq:sign}) through the $e\mu\gamma\gamma$ vertex and the
coefficients $C_{eAAL} = C_{eAAR} =1$.  The numerical results agree
with (\ref{sigmagaga}).

We expect the results for cross sections with cuts in table
\ref{tab:sigmacuts} to be quite realistic, although detector
efficiencies and acceptances have not been allowed for in those
results. However, we expect the rapidity acceptance effects to
be taken into account by the cut $C_3$ in (\ref{eq:cuts2}), and we
also expect the efficiency for photon identification to be no less
than 90\%, so that detector effects should be modest.  The important
exception to this, however, is the background process
(\ref{eq:proc2}), which in table \ref{tab:sigmacuts} seems to
represent one-third of $\sigma_{\gamma\gamma}^{(4)}$, but which must
actually be adjusted for the electron-photon misidentification
probability.  In order to settle this issue, we carried out a detector
simulation using Delphes 3.4 \cite{delphes}.

An $e^-\mu^+$ collider is highly asymmetric, so we assume its detector
to have a correspondingly asymmetric design.  We obtain a simple but
effective asymmetric detector simulation in the lab frame by assuming
that in the CM frame, in which the $e^-\mu^+$ collisions are
forward-backward symmetric on average, the detector possesses the same
capabilities as the symmetric Muon Collider detector implemented in
Delphes 3.4\footnote{Which corresponds to the configuration file
\texttt{cards/delphes\_card\_MuonColliderDet.tcl} in the Delphes
distribution.}. 
We simulate the collisions with \textsc{madgraph} 5 with
a loosened version of the cuts (\ref{eq:cuts2}) in order to adequately
populate the phase space,
\begin{equation}
  \label{eq:cuts3}
  \begin{gathered}
    C'_0=C_0,
  \qquad
  C'_1:\; E_{3}^*, E_{4}^* > 250.0\,\mathrm{GeV},
  \\
  C'_2:\; p_T^\mathrm{tot} = |\vec{p}_{3T}+\vec{p}_{4T}| <
  40.0\,\mathrm{GeV},
  \qquad
  C'_3=\emptyset~.
  \end{gathered}  
\end{equation}
We run Pythia 6 \cite{pythia} on these events for QED showering,
followed by Delphes 3.4 with Muon Collider configuration.  In figure
\ref{fig:distributions} we display normalized differential cross
sections with respect to the CM rapidity $y^*$, the transverse
momentum $p_T$ and the CM energy $E^*$ for the final-state photons in
processes (\ref{eq:sign}), (\ref{eq:proc1}), and the final-state
$e^\pm$ for (\ref{eq:proc2}), at three different collision
energies. Those differential cross sections correspond to
detector-level events generated with the cuts (\ref{eq:cuts3}) at the
parton level. We notice here that the differential cross sections
shown in the figure for the signal process (\ref{eq:sign}) correspond
to the contact interaction (\ref{eq:Q8}) represented by the symbol
$\sigma_{\gamma\gamma}^{(4)}$ in tables \ref{tab:sigmacuts} and
\ref{tab:delphes}.  For the kinematic variables considered in figure
\ref{fig:distributions}, however, identical results would have been
obtained with the interaction (\ref{eq:Q6}). We also point out here
the good agreement of the analytical rapidity distribution
(\ref{rapidityshape}) with the Monte Carlo data in figure
\ref{fig:distributions}, right column.
\begin{figure}
  \includegraphics[scale=0.9]{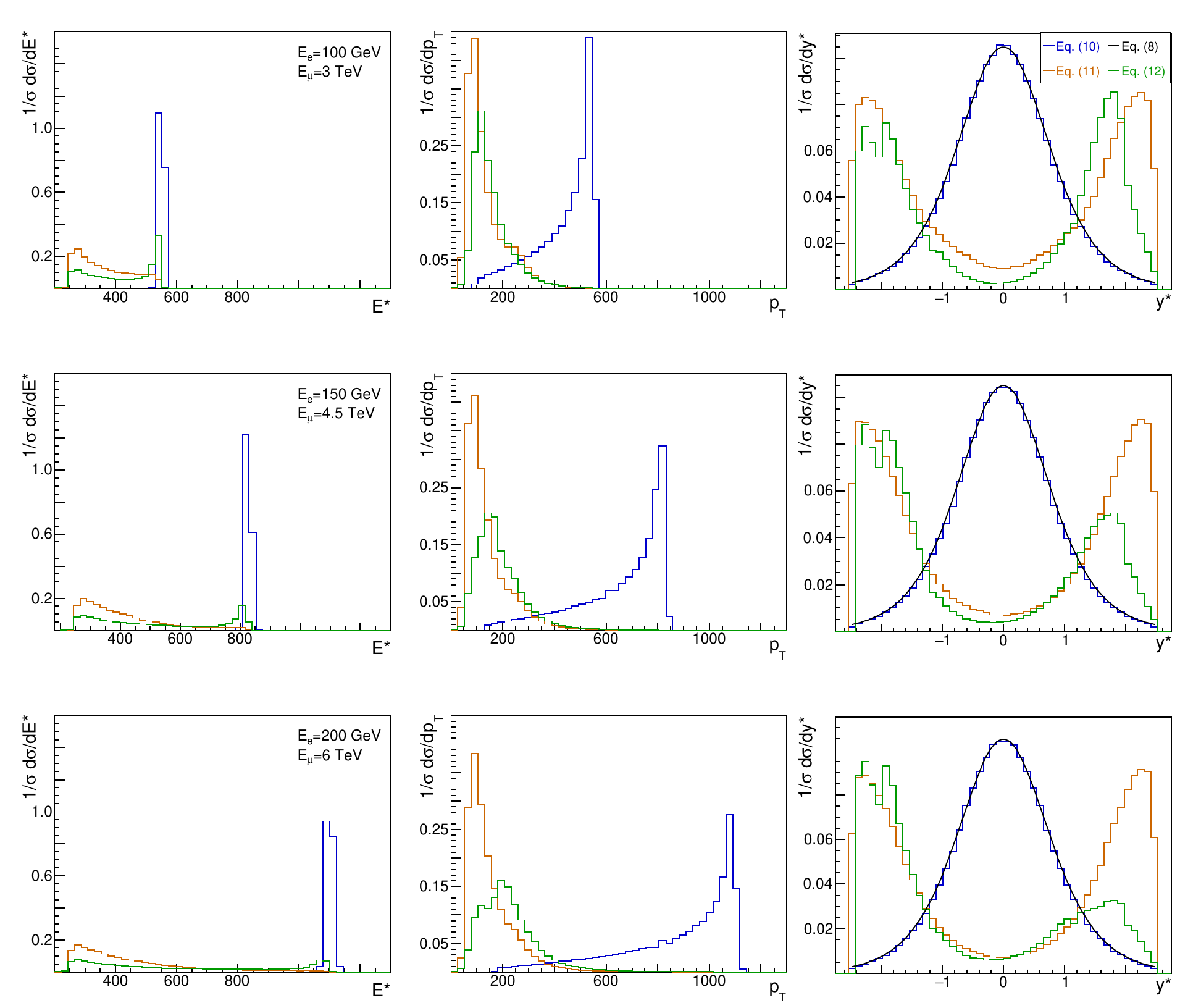}
  \caption{Normalized differential cross sections for final-state
    photons for processes (\ref{eq:sign}), (\ref{eq:proc1}) and
    $e^\pm$ for process (\ref{eq:proc2}), with respect to:
    center-of-mass energy (left column), transverse momentum (center
    column) and center-of-mass rapidity (right column).  The beam
    energies are $(E_e,E_\mu)=$ (100 GeV, 3 TeV) (upper row), (150
    GeV, 4.5 TeV) (middle row), (200 GeV, 6 TeV) (lower row). Blue
    lines: signal process (\ref{eq:sign}), black line: analytical
    expression (\ref{rapidityshape}), orange lines: background process
    (\ref{eq:proc1}), green lines: background process
    (\ref{eq:proc2}). All differential cross section correspond to
    detector-level events with cuts~(\ref{eq:cuts3}).}
  \label{fig:distributions}
\end{figure}

We then apply a preselection cut
\begin{equation}
  \label{eq:cuts4}
  N_\gamma \geq 2, \qquad N_{e^\pm} = 0, 
\end{equation}
to the Delphes events.  Notice that (\ref{eq:cuts4}) implicitly
includes a cut in absolute rapidity $|y^*|<2.5$, corresponding to the
detector rapidity acceptance range. Finally, we apply the cuts
$C_{0\mbox{--}3}$ from (\ref{eq:cuts2}) to the preselected events.
The cross sections obtained for the detector-level events are
summarized in table~\ref{tab:delphes}. We notice here that the cross
section for the process (\ref{eq:sign}) induced by the dim 6 operator
(\ref{eq:Q6}), as displayed by the diagrams in figure \ref{fig:emu}
(a), has an energy-independent cross section at the parton level, but
shows a slight decrease with increasing $\sqrt{s}$ in table
\ref{tab:delphes}. This is due to the fact that the upper
limit on $p_T^\mathrm{tot}$ we are using in (\ref{eq:cuts2}) is fixed.
This effect, however, is more than compensated for by the increase in
the partonic cross section in the case of the signal process induced
by the dim 8 operator (\ref{eq:Q8}), as displayed in
figure \ref{fig:emu} (b), leading to a increasing cross section also
at the detector level.  The cross section for the process
(\ref{eq:proc1}) shows a very modest growth with $\sqrt{s}$ in table
\ref{tab:delphes}, and process (\ref{eq:proc2}) actually decreases at
the highest energy.
\begin{table}[ht!]
  \centering
  \begin{tabular}{|c|c|c|c|c|c|}\hline
$\mathcal{P}_\mu$ & $\mathcal{P}_e$ 
 & $\sigma_{\gamma\gamma}^{(3)}[\mathrm{fb}]$
 & $\sigma_{\gamma\gamma}^{(4)}[\mathrm{fb}]$
 & $\sigma_{\nu\nu\gamma\gamma}[\mathrm{fb}]$
 & $\sigma_{\nu\nu ee}[\mathrm{fb}]$  \\\hline
\multicolumn{6}{|c|}{$E_e=$ 100 GeV, $E_\mu=$ 3 TeV}\\\hline    
    $0.0$ & $0.0$ & $1.72$ & $0.0082$ & $\sim10^{-5}$ & $\sim10^{-5}$ \\\hline    
 $+0.4$ & $+0.8$ & $2.26$ & $0.011$ & $\lesssim10^{-5}$ & $\lesssim10^{-5}$ \\\hline
\multicolumn{6}{|c|}{$E_e=$ 150 GeV, $E_\mu=$ 4.5 TeV}\\\hline    
    $0.0$ & $0.0$ & $1.61$ & $0.039$ & $\sim10^{-4}$ & $\sim10^{-5}$  \\\hline    
 $+0.4$ & $+0.8$ & $2.12$ & $0.051$ & $\lesssim10^{-4}$ & $\lesssim10^{-5}$ \\\hline
\multicolumn{6}{|c|}{$E_e=$ 200 GeV, $E_\mu=$ 6 TeV}\\\hline    
 $0.0$ & $0.0$ & $1.46$ & $0.11$ &  $\sim10^{-4}$ &     $\lesssim10^{-5}$ \\\hline    
  $+0.4$ & $+0.8$ & $1.93$ & $0.15$ & $\lesssim10^{-4}$ & $\lesssim10^{-5}$ \\\hline
  \end{tabular}                               
  \caption{Cross sections in fb for the signal processes
    (\ref{eq:sign}) obtained from the operator $Q_{eA}$, see eq.\
    (\ref{eq:Q6}) and fig.\ \ref{fig:emu} (a), (\ref{eq:sign})
    obtained from the operator $Q_{eAA}$, see eq.\ (\ref{eq:Q8}) and
    fig.\ \ref{fig:emu} (b), and the background processes
    (\ref{eq:proc1}) and (\ref{eq:proc2}). All cross sections, at the
    three energies shown, obtained at the detector-simulation level
    with cuts $C_{0\mbox{--}3}$ from~(\ref{eq:cuts2}).}
  \label{tab:delphes}
\end{table}

Furthemore, comparing the results in table \ref{tab:delphes} at
$(E_e,E_\mu)=$ (100, 3000) GeV with those in table
\ref{tab:sigmacuts}, we see that detector effects result in an
effective efficiency of 84\% for the $\gamma\gamma$ production
processes given in the tables by $\sigma_{\gamma\gamma}^{(3)}$,
$\sigma_{\gamma\gamma}^{(4)}$. We see also that detector efficiencies
reduce the cross section for the $e^-e^+$ background process
(\ref{eq:proc2}) to the same $O(10^{-5})\,\mathrm{fb}$ level as the
$\gamma\gamma$ background (\ref{eq:proc1}).  Completely analogous
results are obtained at the two higher energies considered in table
\ref{tab:delphes}.

\section{Conclusions}
\label{conclusions}

We have obtained individual limits on LFV four fermion operators by
looking at the two-to-two production processes they induce at the
$e^- \mu^+$ collider.  For operators $e\mu f\bar f$ where $f$ is a
second or third family fermion the sensitivity of the collider is of
the same order of magnitude as, and for some operators even somewhat
stronger than, that of the $\mu A \to e A$ conversion in nuclei.
On the other hand, the $e\mu$ collider would have higher sensitity
than the other low energy measurements $\mu \to e{\bar e}e$ and
$\mu \to e \gamma$ even for first family fermions.   In the particular
case of the Wilson coefficients $C_{lq}^{(1,3)}$, for example, the
expected sensitivity at an $e \mu$ collider would be at least as strong
as that of all low-energy measurements, and an order of magnitude
larger than that projected for Drell-Yan processes at the HL-LHC.
This leads us to expect that, given the large number of independent
effective four-fermion operators, the additional information
obtained from the collider will certainly be invaluable.
In the case of $e\mu \to \gamma \gamma$ production the limits from
the collider are significantly more stringent than those from
the low energy processes for the case of the dimension 8
$e\mu \gamma \gamma$ contact operator (\ref{eq:Q8}).

In the SM the $f\bar f$ and $\gamma \gamma$ production involves an
additional pair of neutrinos and this makes the separation of signal
and background straightforward.  We have made a detailed analysis of
background and signal for the case of $\gamma \gamma$ production,
including detector simulation.  We observe that with appropiate cuts
on the photon energies, the photon pair transverse momentum and the
photon rapidities we can dramatically lower the SM background with
very little reduction of the signal.

We point out, finally, that our conclusions are based on the
assumptions of an integrated luminosity of 1 ab$^{-1}$ and beam
energies $(E_e,E_\mu)=$ (100 GeV, 3 TeV), (150 GeV, 4.5 TeV), (200
GeV, 6 TeV), respectively. Clearly, higher luminosities and/or beam
energies would lead to stronger sensitivities to the contact-interaction
effective couplings considered here.

\paragraph*{Acknowledgments}

We are grateful to Georgina Espinoza Gurriz for her assistance with our
computer hardware.  We acknowlegde support from Sistema Nacional
de Investigadores de Conacyt, M\'exico.  We also acknowledge that
a preprint has previously been published at arXiv \cite{lulevin}.

\end{document}